\documentclass[letterpaper, 10 pt, conference]{ieeeconf}  % Comment this line out if you need a4paper

\usepackage{amsmath,amssymb,amsfonts}
\usepackage{graphicx,wrapfig}
\usepackage{subcaption}
\usepackage{makecell}
\usepackage{dsfont}

\usepackage{gensymb}%for the \degree command
\usepackage{color}
\usepackage[colorlinks]{hyperref}

\usepackage{accents}
\newcommand{\ubar}[1]{\underaccent{\bar}{#1}}

\def\rd#1{{\color{red}{#1}}}
\def\pb#1{\footnote{PB:\rd{#1}}}

\IEEEoverridecommandlockouts                              % This command is only needed if 
\usepackage{xparse}

\ExplSyntaxOn
\cs_new:Nn \pb_prop_gset_bykeys:Nn
{
	\prop_clear:N \l__pb_temp_prop
	\keys_set:nn { pb/propbykey } { #2 }
	\prop_gset_eq:NN #1 \l__pb_temp_prop
}
\cs_generate_variant:Nn \pb_prop_gset_bykeys:Nn { c }

\keys_define:nn { pb/propbykey }
{
	unknown .code:n = \prop_put:Nxn \l__pb_temp_prop { \l_keys_key_tl } { #1 }
}

\cs_generate_variant:Nn \prop_put:Nnn { Nx }

\NewDocumentCommand{\setupcollaborator}{mm}
{% #1 = identifier string, #2 = set of key-value pairs
	\prop_new:c { g_collaborator_#1_prop }
	\pb_prop_gset_bykeys:cn { g_collaborator_#1_prop } { #2 }
}

\NewDocumentCommand{\selectcollaborator}{m}
{
	\prop_map_inline:cn { g_collaborator_#1_prop }
	{
		\tl_set:cn { ##1 } { ##2 }
	}
}
\ExplSyntaxOff

\setupcollaborator{PBmac}
{% pb, macbook air or mac mini at work, 
	NAME=Prabir Mac , laptop or mac mini,
	DiCEbibPATH=/Users/pbarooah/Dropbox/Dropbox/DiCElab_bib,
	NRbibPATH = /Users/pbarooah/Dropbox/Dropbox/NSRbib
}
\setupcollaborator{NRlinux}
{% NRlinux is Naren using the linux machine at work
	NAME=NR at lab,
	DiCEbibPATH=../../DiCElab_bib,
	NRbibPATH = ../../NSRbib,
}
\setupcollaborator{NRwindows}
{% NRwindows is Naren using his windows.
	NAME=NR using windows,
	DiCEbibPATH=../../DiCElab_bib,
	NRbibPATH = ../../NSRbib,
}
\setupcollaborator{NGlinux}
{% NGlinux is Ninad using his linux machine at work.
	NAME=NG at Lab,
	DiCEbibPATH=../../../../../DiCElab_bib,
}

\setupcollaborator{NGWindows}
{% NGlinux is Ninad using his linux machine at work.
	NAME=NG using Windows,
	DiCEbibPATH=../../../../../../DiCElab_bib,
}
\selectcollaborator{NGWindows} 
\newif\ifArxivx

\Arxivxtrue
\graphicspath{{./Images/}}

\title{\LARGE \bf
Increasing Energy  Resiliency to Hurricanes with Battery and Rooftop Solar Through Intelligent Control}

\author{Ninad Gaikwad, Naren Srivaths Raman, and Prabir Barooah% <-this % stops a space
\thanks{The authors are with the Department of Mechanical and Aerospace Engineering, University of Florida, Gainesville, Florida 32611, USA.
	{\tt\small ninadgaikwad@ufl.edu, narensraman@ufl.edu, pbarooah@ufl.edu.} The research reported here is partially supported by NSF awards 1646229 and 1934322.}%
}

\begin{document}

\maketitle
\thispagestyle{empty}
\pagestyle{empty}

%%%%%%%%%%%%%%%%%%%%%%%%%%%%%%%%%%%%%%%%%%%%%%%%%%%%%%%%%%%%%%%%%%%%%%%%%%%%%%%%

\begin{abstract}
Rooftop solar photovoltaic (PV) panels together with batteries can provide resiliency to blackouts during natural disasters such as hurricanes. Without intelligent and automated decision making that can trade off conflicting requirements,  a large PV system and a large battery is needed to provide meaningful resiliency. By utilizing the flexibility of various household demands, an intelligent system can ensure that critical loads are serviced longer than a non-intelligent system. As a result a smaller (and thus lower cost) system can provide the same energy resilience that a much larger system will be needed otherwise.

In this paper we propose such an intelligent control system that uses a model predictive control (MPC) architecture. The optimization problem is formulated as a MILP (mixed integer linear program) due to the on/off decisions for the loads. Performance is compared with two rule based controllers, a simple all-or-none controller that mimics what is available now commercially, and a Rule-Based controller that uses the same information that the MPC controller uses. The controllers are tested through simulation on a  PV-battery system chosen carefully for a small single family house in Florida. Simulations are conducted for a one week period during hurricane Irma in 2017. Simulations show that the size of the PV+battery system to provide a certain resiliency performance can be halved by the proposed control system. 
\end{abstract}

%%%%%%%%%%%%%%%%%%%%%%%%%%%%%%%%%%%%%%%%%%%%%%%%%%%%%%%%%%%%%%%%%%%%%%%%%%%%%%%%
\section{Introduction}\label{section:Intro}
Atlantic hurricanes are occurring with increasing frequency~\cite{USGCRPreport:2014}.  Loss of electricity supply for long periods is a common outcome of hurricanes. %A few recent examples include 4.8 millions of utility customers losing electricity in Florida after hurricane Irma, with 1.5 million remaining without electricity for five days or more~\cite{IrmaImpactReport:EIA:2017}, and the months-long blackout in Puerto Rico after hurricane Maria~\cite{GallucciRebuilding:spectrum:2018}, leading to an estimated death toll in the thousands~\cite{KishoreMortality:NEJM:2018}.
Rooftop solar PV panels can provide a resilient energy supply to hurricane-induced blackouts since  the sky is often clear  immediately after the hurricane. Serving the entire household load from an on-site PV+battery system may require a large system depending on the size of the house, driving up cost. %For instance, the average household load in the U.S. is quite high $30.5~kWh/day$~\cite{EIAAnnual:2019}, and a reliable PV+battery backup will be quite expensive.
Even serving only the most critical loads, such as refrigeration and a few lights and fans, during a long blackout in the face of solar generation uncertainty may still require a large PV+battery system.  

We argue that the size---and thus, cost---of the PV+battery system to provide resiliency can be reduced with the help of an intelligent control system. The key is to exploit flexibility in demand as well as in the supply in conjunction with forecasts. Flexibility in demand comes from the fact that, after a disaster not some loads are more critical than others. Among the critical loads that need to be served during a blackout, refrigeration for food and medicine is the most important~\cite{CdcKeep:2019}. Next comes lights, and then fans. Fans can serve as temporary replacements for air conditioners to provide thermal comfort, and are much less energy intensive than  air conditioners. An intelligent controller can prioritize the refrigerator demand over light and fan demand, and turn off all other loads. Flexibility in supply comes from the fact that the charging rate of batteries is variable; a battery can be fast charged to prepare for a forecasted low solar irradiance event, though at some cost to the  battery's health.

Thus, by using forecast of solar generation and household demand, an intelligent decision maker can operate the equipment (battery, primary loads and secondary loads) to ensure that a higher energy resiliency is obtained. Without such intelligence, a much larger system will be needed to deliver the same energy resilience.

This paper presents such an intelligent control system for a home in which the critical loads are a refrigerator, a few lights and fans, and a small PV+battery system. Among the critical loads that need to be served with on-site energy during an outage, the refrigerator is deemed \emph{primary} load while the combined load of the lights and fans is called the \emph{secondary} load. It is more important to serve the primary than the secondary load. The goal of the control system is to keep the refrigerator temperature within a band while serving the secondary demand as much as possible. Although there are many more electrical loads in a typical household, others are not critical for health and well being after a disaster. For instance, though many homes use electric cook-tops for cooking in Florida, people often use outdoor gas grills to cook food after hurricanes~\cite{NorrisHow:2017,OttensteinGet:2019}.

A model predictive control (MPC) architecture is used that uses available measurements and forecasts to make optimal decisions in real time. The optimization problem is formulated as a MILP (mixed integer linear program); the integer valued variables are for the on/off status of power supply to the  primary and secondary loads (refrigerator and the aggregate demand of lights and fans). Performance is compared with two rule based controllers, a simple \emph{baseline} controller that mimics what is available now commercially, and another controller -- that we call Rule-Based controller -- that uses the same information that the MPC controller uses. Simulations show that the proposed MPC controller is able to service the primary load (refrigerator) throughout the simulation period (7 days after hurricane Irma in 2017) while the baseline controller is unable to do so for several hours each day. In addition, it is able to service the secondary load a little more than the baseline. We measure \emph{primary resiliency performance} of a control system as the average daily duration that the system is able to meet demand from the primary load. A simulation based study indicates that to meet a specific primary resiliency performance, the cost of the PV+battery system needed by the baseline controller is \emph{twice} that of that needed by the MPC controller. The cost of energy resiliency can therefore be halved by the MPC control system. 

All three controllers utilizes the same hardware, and any modification to the household electrical wiring to implement them is the same. The difference is in the sophistication of their decision making and available sensing. The MPC and Rule-Based controller use two more sensors (refrigerator temperature and house temperature) and uses solar irradiance forecasts compared to the baseline controller. 

%A dynamic model of the refrigerator is used to decide its compressor on/off status so that its temperature stays within an allowable band. Since fast charging of the battery degrades life, the controller tries to use normal charging as much as possible, using fast charging only when absolutely necessary.

In this work we only focus on the control algorithm for post-disaster scenario in which grid supply has been lost. It is assumed that when grid supply is restored, the software will switch to a ``normal operating'' mode. The normal operating mode may also be a sophisticated controller that seeks to, for instance, minimize the utility bill of the consumer by controlling the PV+battery system. There is a plethora of work in that direction; see~\cite{DiEvent:2012},~\cite{AnvariOptimal:2014},~\cite{BrahmanOptimal:2015},~\cite{MarzbandOptimal:2017} and~\cite{SanjariAnalytical:2017}. Therefore we do not consider that problem here. Works on controlling the PV+battery system to maximize resiliency performance in a post-disaster scenario, the focus of this paper, is extremely limited. To the best of our knowledge, only~\cite{PrinceResilience:2019} and~\cite{PrinceStochastic:2020} consider the problem of operation for resiliency. However, both \cite{PrinceResilience:2019} and~\cite{PrinceStochastic:2020} ignore the mixed-integer nature of the optimization problem, and ignores the capability of a battery to vary charging rate which can be exploited during contingency situations like power outage.

% The PV+battery system parameters are chosen according to existing design guidelines for standalone PV+battery systems as given in~\cite{MastersRenewable:2013}. The primary and secondary loads are chosen to be representative of what one might encounter in a typical home in Florida. For comparison, we also simulate a baseline controller that is representative of the existing commercial systems one can install today. It energizes both the primary and secondary loads' circuits if it estimates that there is enough energy available from the PV+battery system to service the combined demand.

A preliminary version of this work was presented in~\cite{GaikwadSmartCCTA:2020}. We have made several improvements in this paper. (i) We introduce a Rule-Based controller in this paper that uses the same sensors and forecasts, and even the dynamic models, that MPC uses to make decisions. The purpose is to check whether the performance of the MPC controller is achievable with a simpler controller that does not need real-time optimization. Results indicate it is unlikely. (ii) The MPC controller in~\cite{GaikwadSmartCCTA:2020} needed a dynamic model of the house (indoor) temperature - apart from a sensor - for prediction. Obtaining parameters of such models is challenging~\cite{GaikwadSmartCCTA:2020}. The MPC controller proposed here uses a simpler parametric model of the indoor temperature that only needs a temperature sensor and historical data on outdoor temperature for predicting indoor temperature. (iii) The controller in~\cite{GaikwadSmartCCTA:2020} assumed the ability  to turn the refrigerator on and off. While turning off is easy, forcing the compressor on is difficult with an existing refrigerator without expensive retrofit. The controller in this paper assumes only the ability to interrupt power supply to the refrigerator, not the ability to force it on. Such actuation is far cheaper in practice, with a wireless controlled smart switch, than the ability to command the refrigerator compressor to turn on. (iv) We have reported here several sensitivity studies, while the simulations in~\cite{GaikwadSmartCCTA:2020} were only for one specific set of parameters.

The rest of the paper is organized as follows: the system description and mathematical models of the system components are described in Section~\ref{section:PlantModel}. The formulation of MPC controller, and the description of Rule-Based and baseline controllers are provided in Section~\ref{section:ControlAlgorithms}. The simulation setup, computation and simulation parameters discussed in Section~\ref{section:SRD}. The results of the simulation study are presented and discussed in Section~\ref{section:ResultsDiscussions}. Finally, the main conclusions are provided in Section~\ref{section:Conclusions}.

\section{System Description}\label{section:PlantModel}
\begin{figure}[t]
	\centering
	\includegraphics[scale=0.3]{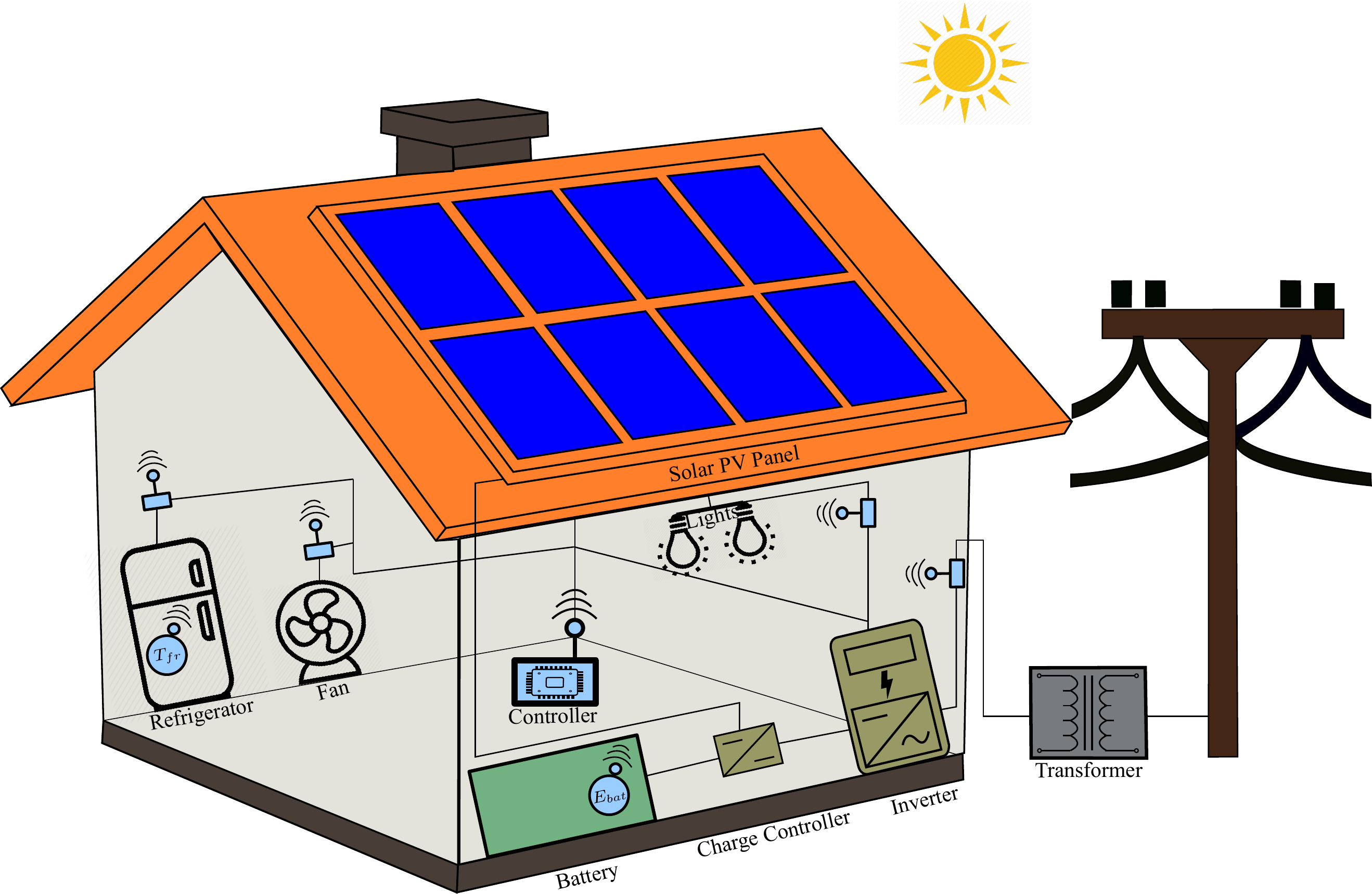}
	\caption{Hardware involved in the intelligent control system.}
	\label{fig:PhysicalPlant_New}
\end{figure}

Figure~\ref{fig:PhysicalPlant_New} shows the schematic of a house with solar PV panels, a battery energy storage system, a  primary load (refrigerator), and secondary loads (lights+fans). An intelligent controller needs to trade-off primary-secondary demand and battery life to provide 'best' service possible during an extended outage within certain constraints. The primary goal is to maintain the refrigerator temperature within the prescribed limits, and a secondary goal is to service the secondary load during times that are pre-decided by the occupants.

In order to achieve these goals, the intelligent controller needs to control the following: (i) power supply to the refrigerator, (ii) on/off state of the secondary load (aggregate of lights and fans),  (iii) charging/discharging state of the battery, and (iv) when charging the battery, the charging mode of  the battery. The battery has two charging modes: normal and fast. The refrigerator power supply is actuated using a smart switch. The intelligent controller can control only the power supply to the smart switch to which the refrigerator is connected, but not the actual on-off of the refrigerator compressor which is controlled by the refrigerator thermostat. 
%\begin{align}
%u_{fr}(k)= 
%\begin{cases}
%1,& \text{if }  T_{fr}(k) \geq \bar T_{fr} \text{ AND } \\
%& u_{frc}(k)=1\\
%0,& \text{if }  T_{fr}(k) \leq \ubar T_{fr} \text{ OR }\\
%& u_{frc}(k)=0\\
%u_{fr}(k-1),& \text{otherwise, }
%\end{cases} \label{eq:FridgeOnOff_MILP}
%\end{align}. 
The secondary load on-off is actuated through a secondary circuit coupled with a smart switch. The battery actuation can be done through the smart charge controller. Moreover, the intelligent controller requires the following additional information: (i) forecasted irradiance, (ii) measured house temperature, (iii) measured internal refrigerator temperature, and (iv) measured battery state of charge. Forecasts for the irradiance are obtained from a provider over the internet. It is assumed that the controller stores forecasts received over at least a week, which can be used to fit a simple time series model to provide local forecasts of the irradiance in case the internet is down during the extended outage. The house and internal refrigerator temperatures are sensed using smart temperature sensors which communicate with the controller over the air, while the battery state of charge can be obtained over the air from a smart charge controller. 

Mathematical models of each of these components are described in the subsections below. Time is discrete, with $k=0,1,2,\dots$ denoting the time index and $\Delta t_s$ denoting the interval (hours or minutes) between $k$ and $k+1$. In the sequel,  $E(k)$ ($Wh$) will denote the energy consumed/generated during the time interval between time indices $k$ and $k+1$, with the subscript specifying the source or consumer of the energy. The dependence on $k$ will be often omitted, e.g., we will say $x$ instead of $x(k)$. 

The plant used for closed loop simulations consists of the dynamic models: Battery energy storage system model, refrigerator thermal model and the energy consumption models of the household electric loads, presented in Section~\ref{section:ControlAlgorithms}, and the interactions between the PV panels, the battery, and the loads (both primary and secondary). These interactions are represented mathematically using the following equations:
\begin{align}
%&E_{pv}(k) = E_{pv}^{u}(k)+E_{pv}^{un}(k), \label{eq:PlantModel_1} \\
&E_{hl}(k) = \dfrac{u_{fr}(k)E_{fr}(k)+u_{s}(k)E_{s}(k)}{\eta_{inv}}, \label{eq:PlantModel_2} \\
%&E_{pv}^{u}(k) = E_{hl}(k)+E^{c}_{bat}(k), \label{eq:PlantModel_3} \\
&E^{c}_{bat}(k) = c(k) \; min\big\{E_{pv}(k)-E_{hl}(k), \nonumber \\ 
& \quad \quad \quad \quad \quad \bar E_{bat}-E_{bat}(k),  x_{bat}(k) \; \bar E^{c}_{bat} \big\}, \label{eq:PlantModel_5} \\
&E^{dc}_{bat}(k) = d(k) \; min\big\{E_{hl}(k)-E_{pv}(k), \nonumber \\  
& \quad \quad \quad \quad \quad E_{bat}(k)-\ubar E_{bat}, \bar E^{dc}_{bat}\big\}. \label{eq:PlantModel_6}
\end{align}
Eq.~\eqref{eq:PlantModel_2} shows that the total house load ($E_{hl}$) is composed of the energy used by the refrigerator ($E_{fr}$) and the secondary loads ($E_{s}$), where $\eta_{inv}$ is the inverter efficiency, $u_{fr}$ is the refrigerator power supply on-off control command, and $u_{s}$ is the on-off command signal for the secondary load. 
Eq.~\eqref{eq:PlantModel_5} shows that the battery charging energy ($E^{c}_{bat}$) is such that it never charges beyond the maximum limit ($\bar E_{bat}$), where $E_{bat}$ is the battery energy level, $x_{bat}$ is either 1 or 2 (1 - normal charging and 2 - fast charging), $\bar E^{c}_{bat}$ is the maximum battery charging energy, $c$ is the battery charging control command ($c=1$ for charging, $c=0$ for not charging), and $E_{pv}$ is maximum available energy that can be produced by the PV panels. Eq.~\eqref{eq:PlantModel_6} shows that the battery never discharges below the minimum limit ($\ubar E_{bat}$), where $E^{dc}_{bat}$ is the battery discharging energy, $\bar E^{dc}_{bat}$ is the maximum battery discharging energy, and $d$ is the battery discharging control command ($d=1$ for discharging, $d=0$ for not discharging). 

\section{Control Algorithms}\label{section:ControlAlgorithms}
%\begin{figure*}[t]
%	\centering
%	\includegraphics[scale=0.85]{MILPOnly_ControlSchematic_New5.pdf}
%	\caption{Schematic of closed loop operation of the MPC control system.}
%	\label{fig:MILP_Schematic}
%\end{figure*}
\subsection{Model Predictive Control (MPC)}\label{subsection:MPC}
The control decisions are computed at discrete time steps $k = 1,2 \dots N$  with $\Delta t_{s}$ as the sampling period, and $N$ is the total number of time steps in the planning/prediction horizon. The decision variables for the optimization problem elemental to the MPC controller are as follows: the states of the process $x(k) = [E_{bat}(k),T_{fr}(k)]^{T}$, where $T_{fr}(k)$ is the refrigerator internal temperature; the control commands $u(k) = [\Gamma(k),u_{fr}(k),u_{s}(k)]^{T}$, where $\Gamma(k)$ is the fraction of the normal battery charging energy; the internal variables $v(k) = [g(k),\zeta_{fr}(k)]^{T}$, where $g(k)$ is the energy produced by the PV panels, and $\zeta_{fr}(k)$ is a slack variable for refrigerator temperature to ensure feasibility. The exogenous inputs whose predictions are assumed to be known for the $N$ time steps $w(k) = [E_{pv}(k),T_{house}(k),E_{s}(k)]^{T}$, where $E_{pv}(k)$ is the available energy from the PV panels computed using \eqref{eq:PV_Model}, $E_{s}(k)$ is the total secondary load and is assumed to be known ahead of time and can be decided by the occupants, and $T_{house}(k)$ is the internal house temperature which is estimated using a simple parametric model described in \eqref{eq:HouseThermalParametricModel}. Hence, the complete decision vector for the optimization problem is given as $[X,U,V]^{T}$, where $X:=[x(k+1), \dots ,x(k+N)]^{T}$, $U:=[u(k), \dots , u(k+N-1)]^{T}$ and $V:=[v(k), \dots , v(k+N-1)]^{T}$.

%A constrained optimization problem is solved, to generate control commands for N time steps, which tries to keep the refrigerator temperature in the prescribed temperature range, maximize the battery state of charge, minimize the health degradation of the battery, and maximize the operation of the secondary loads. This objective is achieved subject to constraints on the refrigerator and battery model dynamics, energy balance equation, battery state of charge constraints, and the constraints on charging and discharging rate of the battery. The optimization problem at any time index $j$ is given mathematically as follows:
A constrained optimization problem is solved, to generate control commands for N time steps, which tries to maximize the primary and secondary loads, while minimizing the battery degradation and maximizing system life. This objective is achieved subject to constraints on dynamics, energy balance, states and control commands. The optimization problem at any time index $j$ is:
\begin{subequations}
\begin{align} 
	\min_{U} & \sum\limits^{j+N-1}_{k=j} \bigg[\lambda_{1}(N-k)\zeta_{fr}(k) - \lambda_{2}E_{bat}(k) + \\ 
	&\quad \quad \quad \quad \lambda_{3}\Gamma(k) - \lambda_{4}(N-k)u_{s}(k) \bigg],  \label{eq:CostFunction} \\
%\end{aligned}
&\text{subject to:} \nonumber \\ 
%\begin{align}
	&T_{fr}(k+1)=AT_{fr}(k) + Bu_{fr}(k) Q_{fr} + D T_{house}(k), \label{eq:Fridge_eqcon_1} \\
	&E_{bat}(k+1) = E_{bat}(k) + \Gamma(k)  \eta_{bat}^{c,dc,con} \bar E_{bat}^{c}, \label{eq:Battery_eqcon_1}\\
	&u_{fr}(k) E_{fr} + \Gamma(k) \bar E_{bat}^{c} + u_{s}(k)E_{s}(k)=  g(k), \label{eq:EnergyBalance_eqcon_1} \\
	&\ubar T_{fr}\leq T_{fr}(k)\leq \bar T_{fr} + \zeta_{fr}(k), \label{eq:Fridge_ineq_1} \\
	&\zeta_{fr}(k)\geq 0, \label{eq:Slack_ineq_1} \\
	&\ubar E_{bat}\leq E_{bat}(k)\leq \bar E_{bat}, \label{eq:Battery_ineq_1} \\
	&\ubar u_{s} \leq u_{s}(k) \leq \bar u_{s}(k), \label{eq:NCControl_ineq_1} \\
	& \ubar\Gamma \leq \Gamma(k) \leq \bar\Gamma, \label{eq:Battery_ineq_2} \\
    &0 \leq g(k) \leq E_{pv}(k), \label{eq:PV_ineq_1}
\end{align}
\end{subequations}
and only the $j^{th}$ control is implemented. The process is repeated at $j+1, \cdots$.

The cost function in (\ref{eq:CostFunction}) consists of four terms. The first term, $\lambda_{1}(N-k)\zeta_{fr}(k)$, penalizes the refrigerator temperature slack variable which in turn tries to keep the refrigerator internal temperature within prescribed bounds. The time varying weighing factor $N-k$ puts a higher penalty on the temperature violations at earlier times and less weight on violations during later times of the planning horizon. The second term, $- \lambda_{2}E_{bat}(k)$, penalizes a low state of charge which helps in extending the life-time of the system. The third term, $\lambda_{3}\Gamma(k)$, penalizes fast charging, as it is undesirable since it degrades battery life. $\Gamma$ models the fraction of the charging/discharging energy of the battery which are continuous and variable; moreover, it depends on the amount of energy available from the PV panels during charging, and the amount of the load demand to be supplied during discharging. The fourth term, $- \lambda_{4}(N-k)u_{s}(k)$, along with the inequality constraint \eqref{eq:NCControl_ineq_1} maximizes the operation of the secondary loads when desired. The reason for the time varying weight in this term is similar to that in the first term. The parameters $\lambda_{1}, \lambda_{2}, \lambda_{3}$, and $\lambda_{4}$ are designer specified wights. 

The equality constraint (\ref{eq:Fridge_eqcon_1}) is due to the thermal dynamics of the refrigerator. This is the discretized form of the continuous time refrigerator thermal dynamic model presented in~\cite{CostanzoGrey:2013}. Where, $Q_{fr}~(W)$ is the thermal power rejected by the refrigerator to the ambient when the compressor is on, and $T_{house}~(^{\circ}C)$ is the average internal house temperature. $T_{house}$ is computed using a data driven parametric model given in (\ref{eq:HouseThermalParametricModel}) rather than relying on a house thermal dynamics model, which increases the complexity of implementation. In (\ref{eq:HouseThermalParametricModel}),  $T_{meas}$ is the the internal house temperature measured using a sensor, and $T_{hist}$ is the averaged historical ambient temperature of the location in which the house is situated obtained from a credible weather data repository. The averaged historical ambient temperature is created as a single daily file for each day of the year.
\begin{align}
T_{house}(k)= & 
\begin{cases}
T_{meas} \; , & \text{if } k = 1 \\
\left(T_{meas}-T_{hist}(k)\right)\\
+T_{hist}(k) \; , & \text{if } k \in {2,3,..,N} \\
\end{cases} \label{eq:HouseThermalParametricModel}
\end{align}
%Also, the refrigerator on-off control command affects the refrigerator dead-band controller in (\ref{eq:FridgeOnOff_Local}) in the manner illustrated in (\ref{eq:FridgeOnOff_MILP}), as the MPC controller can control only the on-off of the smart switch to which the refrigerator is connected, but not the actual on-off of the refrigerator compressor which is controlled by the local dead-band controller. 
%\begin{align}
%u_{fr}(k)= 
%\begin{cases}
%1,& \text{if }  T_{fr}(k) \geq \bar T_{fr} \text{ AND } \\
%& u_{frc}(k)=1\\
%0,& \text{if }  T_{fr}(k) \leq \ubar T_{fr} \text{ OR }\\
%& u_{frc}(k)=0\\
%u_{fr}(k-1),& \text{otherwise, }
%\end{cases} \label{eq:FridgeOnOff_MILP}
%\end{align}
In addition, $Q_{fr}=COP \; P_{fr}^{rated}$, where $COP$ is the coefficient of performance, and $P_{fr}^{rated}$ is the rated power consumption of the refrigerator. $A$, $B$, and $D$ are the discrete time equivalents of the continuous time model given in \cite{CostanzoGrey:2013}, which are functions of the thermal resistance ($R_{fr}$) and thermal capacitance ($C_{fr}$) of the refrigerator.
%and they are given by:
%\begin{align*}
%&A=e^{A_{c}\Delta T_{s}}, \quad B=\frac{1}{A_{c}}\left( e^{A_{c}\Delta T_{s}} - 1\right) B_c,\\
%&D=\frac{1}{A_{c}}\left( e^{A_{c}\Delta T_{s}} - 1\right) D_c, 
%\end{align*} 
%where $A_{c}$, $B_{c}$, and $D_{c}$ are the continuous time constants of the model given as follows:
%\begin{align*}
%&A_{c}=\dfrac{-1}{C_{fr}R_{fr}}, \quad B_{c}=\dfrac{-1}{C_{fr}}, \quad D_{c}=\dfrac{1}{C_{fr}R_{fr}}, 
%\end{align*}
%where $R_{fr}~(^{\circ}C/W)$ and  $C_{fr}~(J/^{\circ}C)$ are the thermal resistance and thermal capacitance of the refrigerator respectively. 
The plant utilizes the exact same model for refrigerator thermal dynamics, hence there is no plat-model mismatch.

The equality constraint \eqref{eq:Battery_eqcon_1} is due to the battery energy dynamics, where $\bar E_{bat}^{c}$ is the maximum battery charging and discharging energy (which are assumed to be equal for modeling simplicity) in the normal mode, and $\eta_{bat}^{c,dc,con}$ is the charging-discharging efficiency of battery used in the controller. This battery dynamics differ from plant battery dynamics given in eq.~\eqref{eq:Battery_Model_1} as it models the battery charging and discharging energies with a single continuous variable ($\Gamma$), hence there is no plat-model mismatch. The battery storage system for the plant is simply modeled as a bucket of energy and its dynamics are as follows:
\begin{align} 
E_{bat}(k+1) = E_{bat}(k)+\eta^{c}_{bat}E^{c}_{bat}(k)-\dfrac{E^{dc}_{bat}(k)}{\eta^{dc}_{bat}}, \label{eq:Battery_Model_1}
\end{align} 
%where $E_{bat}$ ($Wh$) is the battery energy level, $E_{bat}^{c}~(Wh)$ is the energy absorbed (charging energy) by the battery to charge, and $E^{bat}_{dc}~(Wh)$ is the energy supplied (discharging energy) by the battery. Also,
where, $\eta^{c}_{bat}$ and $\eta^{dc}_{bat}$ are the battery charging efficiency and the battery discharging efficiency, respectively. The battery energy level is bounded between the minimum ($\ubar E_{bat}$) and maximum ($\bar E_{bat}$) battery energy limits, i.e. $E_{bat}\in[\ubar E_{bat},\bar E_{bat}]$. The charging and discharging energies for the battery are constrained by the maximum charging and discharging energies as $E_{bat}^{c} \in [0,\bar E^{c}_{bat}]$ and $E_{bat}^{dc} \in [0,\bar E^{dc}_{bat}]$.
%where $\bar E^{c}_{bat}~(Wh)$ and $\bar E^{dc}_{bat}~(Wh)$ are the maximum energies that the battery can absorb and supply during $\Delta t_s$, respectively. 

The equality constraint \eqref{eq:EnergyBalance_eqcon_1} is the energy balance equation, where the electrical energy consumed by the primary (refrigerator) and secondary (lights and fans) loads is simply the integral of their rated powers times the number of individual load units. %The energy consumed by the secondary loads, $E_{s}~(Wh)$ is given as follows:
%\begin{align}\label{eq:Total_NC_Load}
%E_{s}(k) = E_{l}(k) + E_{f}(k).
%\end{align}   

The inequality constraint \eqref{eq:Fridge_ineq_1} is to maintain the refrigerator temperature within the lower ($\ubar T_{fr}$) and upper ($\bar T_{fr}$) temperature limits. The inequality constraint \eqref{eq:Slack_ineq_1} is present to not allow the refrigerator temperature slack to become negative. The inequality constraint \eqref{eq:Battery_ineq_1} bounds the battery energy between the minimum ($\ubar E_{bat}$) and maximum ($\bar E_{bat}$) battery energy limits. The inequality constraint \eqref{eq:NCControl_ineq_1} is present to force the secondary load control command to be zero when secondary loads are not desired to be turned on by the occupants, where $\ubar u_{s}$ and $\bar u_{s}$ are the lower and upper bound on $u_{s}$ respectively, and are defined as follows:
\begin{align}
\bar u_{s}(k)= & 
\begin{cases}
1 \; , & \text{if } E_{s}(k) > 0 \\
0 \; , & \text{if } E_{s}(k) = 0 \\
\end{cases}\\
\ubar u_{s}(k)= & 0, \forall\;\; k = 1,2,\dots,N.
\end{align}
The inequality constraint (\ref{eq:Battery_ineq_2}) bounds the fraction of battery charging/discharging energy ($\Gamma$) between a minimum ($\ubar\Gamma=-1$) and maximum ($\bar \Gamma=2$) value. For negative values the battery discharges; whereas for positive values till 1 it charges in normal mode with a battery charging energy of $\bar E_{bat}^{c}$ as the maximum, for values above 1 it charges in fast mode with twice the normal battery charging energy of $2 \times \bar E_{bat}^{c}$ as the maximum. 

The inequality constraint \eqref{eq:PV_ineq_1} bounds the energy produced by the PV panels such that it cannot be negative and is always less than or equal to the maximum available PV energy ($E_{pv}$). $E_{pv}$ is estimated using the following:
%\begin{align} 
%E_{pv}(k) & =  N_{pv} \; P_{pv}^{rated} \; \left( \dfrac{G(k)} {G_{std}} \right)\times \;     \nonumber \\
%& \quad\left( 1 + \frac{\gamma}{100} \;\big(T_{m}(k)-T_{std}\big)  \right) \; \Delta T_{s}, \label{eq:PV_Model} 
%\end{align} 
\begin{align} 
E_{pv}(k) & =  N_{pv} \; P_{pv}^{rated} \; \left( \dfrac{G(k)} {G_{std}} \right)\times  \; \Delta t_{s}, \label{eq:PV_Model} 
\end{align}
where $N_{pv}$ is the number of PV panels, $P_{pv}^{rated}$ ($W$) is the rated power output of PV module, $G_{std}$ ($W/m^{2}$) is the solar irradiance standard test condition respectively, and $G$ ($W/m^{2}$) is the current solar irradiance. Eq.~\eqref{eq:PV_Model} is a modified version of that used in~\cite{TanakaOptimal:2012}.
%; we use $P_{pv}^{rated}$ instead of PV conversion efficiency and array area which is equivalent. 

%, and $T_{m}$ instead of ambient temperature as it is a more accurate way of computing effect of ambient temperature on PV power~\cite{MastersRenewable:2013}.
%The module temperature can be estimated from the ambient air temperature ($T_{am}$ in $^{\circ}C$) and wind speed ($W_s$ in $m/s$) using Faiman's formula~\cite{FaimanAssessing:2008} as follows:
%\begin{align} 
%T_{m}(k)=T_{am}(k)+\dfrac{G(k)}{U_{0}+U_{1}+W_s(k)}, \label{eq:Faiman_Model} 
%\end{align} 
%where $U_{0}$ ($W/m^{2}K$) is the constant heat transfer component and $U_{1}$ ($W/m^{2}K$) is the convective heat transfer component.

The control variables $u_{fr}$ and $u_{s}$ are modeled as binary integer variables, taking values in $\{1,0\}$ to turn the loads on and off respectively. This binary nature of the control commands makes this problem a Mixed Integer Linear Program (MILP).

%Figure~\ref{fig:MILP_Schematic} shows the control architecture for the MPC controller. 
The implementation of the control commands computed by the MPC Controller is as follows. The control commands $ u_{fr}$ and $u_{s}$ are directly applied to the plant, turning the refrigerator and the secondary loads on and off depending on whether $u_{fr}$ and $u_{s}$ are 1 and 0 respectively. However, $\Gamma$ is converted into appropriate discrete decisions, $c$, $d$ and $x_{bat}$, which are then applied to the charge controller in the following manner:
\begin{align}
	c(k)= & 
	\begin{cases}
		1 \; , & \text{if } \Gamma(k) > 0 \\
		0 \; , & \text{if } \Gamma(k) \leq 0  \\
	\end{cases} \label{eq:Battery1} \\
	d(k)= &
	\begin{cases}
		1 \; , & \text{if } \Gamma(k) < 0 \\
		0 \; , & \text{if } \Gamma(k) \geq 0  \\
	\end{cases} \label{eq:Battery2} \\
	x_{bat}(k)= &
	\begin{cases}
		1 \; , & \text{if } 0 < \Gamma(k) \leq 1 \\
		2 \; , & \text{if } 1 < \Gamma(k) \leq 2 \\
		0 \; , & \text{otherwise } .   \\
	\end{cases} \label{eq:Battery3}
\end{align}

\subsection{Baseline Controller}\label{subsection:Baseline}
%\begin{figure}[htpb]
%	\centering
%	\includegraphics[scale=0.6]{Baseline_ControlSchematic_New3.pdf}
%	\caption{Schematic of Baseline Controller.}
%	\label{fig:Baseline_Schematic}
%\end{figure}
The baseline controller is what is commercially available now when one installs a PV+battery backup system. It simply  supplies power to the primary and secondary loads as long as there is enough supply from the PV and/or the battery.  Otherwise it turns off supply to the loads. The thermostat controls the on-off ($u_{fr}$) of the refrigerator compressor. It turns the compressor 'on' when the refrigerator internal temperature ($T_{fr}$) goes above the maximum limit ($\bar T_{fr}$), turns it 'off' when $T_{fr}$ goes below the minimum limit ($\ubar T_{fr}$), and uses the previous control command otherwise. Note that turning the fridge on is only possible if there is power supply, otherwise a ``on'' decision has no effect.
%\begin{align}
%	u_{fr}(k)= 
%	\begin{cases}
%		1,& \text{if }  T_{fr}(k) \geq \bar T_{fr}\\
%		0,& \text{if }  T_{fr}(k) \leq \ubar T_{fr}\\
%		u_{fr}(k-1),& \text{otherwise, }
%	\end{cases} \label{eq:FridgeOnOff_Local}
%\end{align}
%where $\bar T_{fr}$ and $\ubar T_{fr}$ are the maximum and minimum refrigerator temperature limits respectively. 
The charging ($c$) and discharging ($d$) of the battery is controlled by the charge controller as: charging is turned 'on' ($c=1$) when there is excess energy available from PV after servicing both primary and secondary load demands and is turned 'off' ($c=0$) otherwise; discharging is turned 'on' ($d=1$) when the energy available from PV is not sufficient to satisfy the primary and secondary load demands and is turned 'off' ($d=0$) otherwise.
%\begin{align}
%c(k)= 
%\begin{cases}
%1,& \text{if } E_{pv}(k) >  E_{hl}(k) \\
%0,& \text{otherwise }  \\
%\end{cases}\\
%d(k)= 
%\begin{cases}
%1,& \text{if } E_{pv}(k) <  E_{hl}(k) \\
%0,& \text{otherwise. } \\
%\end{cases}
%\end{align}
However, the amount of charging energy depends on the surplus energy production from the PV panels after the house loads have been serviced and the battery energy level as given in \eqref{eq:PlantModel_5}; and the amount of discharge energy depends on the house load energy not served by the PV panels and the battery energy level as given in \eqref{eq:PlantModel_6}. The baseline controller uses only the normal charging mode; it does not employ fast charging.

\subsection{Rule-Based Controller\label{subsection:IntelligentBaseline}}
%The Rule-Based controller is an attempt to try and duplicate the MPC controller using rule based logic, in order to show that the control problem at hand indeed requires a MILP based MPC controller for optimal results. It consists of the following three sub-units which compute the control commands at every time step;
The Rule-Based controller is an attempt to try and duplicate the MPC controller using rule based logic avoiding the sophisticated optimization approach. It consists of the following three sub-units which compute the control commands at every time step;
%The Rule-Based controller is an attempt to try and duplicate the MPC controller using rule based logic, in order to show that the control problem at hand indeed requires a MILP based MPC controller for optimal results. This controller does not control the refrigerator, rather the local dead-band controller controls the refrigerator, and having the information exactly equivalent to the MPC controller it makes decisions for the battery and secondary load. It consists of the following three sub-units;
%\begin{figure*}[!t]
%	\centering
%	\includegraphics[scale=0.85]{IntelligentBaseline_ControlSchematic_New4.pdf}
%	\caption{Schematic of closed loop operation of the Rule-Based Controller.}
%	\label{fig:IntelligentBaseline_Schematic}
%\end{figure*}

\subsubsection{N Time Steps Simulation Model}\label{subsubsection:IntelligentBaseline1}
%It consists of the exact plant model (no plant-model mismatch), which utilizes the rule based baseline controller with fast charging capability to simulate $N$ steps of the plant; while utilizing the same information as the MPC controller, see Eq.~\eqref{eq:IntelligentBaseline_1}. In the process it computes the total energy generation from the pv and battery ($E_{\text{\emph{total generated}}}$) and the total house load demand serviced ($E_{\text{\emph{total demand}}}$), with which the energy mismatch ($E_{Mis}$) is computed as in Eq.~\eqref{eq:IntelligentBaseline_2}. $E_{Mis}$ is utilized to make the control decision for secondary load as described in Section~ \ref{subsubsection:IntelligentBaseline2}. 
%\begin{figure}[!htpb]
%	\centering
%	\includegraphics[scale=0.85]{IntelligentBaseline_OneDaySimulationSchematic_New5.pdf}
%	\caption{Schematic of N Time Steps Simulation Model.}
%	\label{fig:IntelligentBaseline_Schematic1}
%\end{figure}
%\begin{align}
%&\left[E_{\text{\emph{total generated}}}, E_{\text{\emph{total demand}}}\right]^{T}=f(E_{s}(k:k+N), \nonumber  \label{eq:IntelligentBaseline_1} \\
%&T_{house}(k:k+N),G(k:k+N),u_{fr}(k),T_{fr}(k-1), \nonumber \\
%&E_{bat}(k-1)), \\
%&E_{Mis} = E_{\text{\emph{total generated}}} - E_{\text{\emph{total demand}}} , \label{eq:IntelligentBaseline_2}
%\end{align}
It consists of the plant model, which utilizes the rule based baseline controller with fast charging capability, see Section~ \ref{subsubsection:IntelligentBaseline3} . The fast charging is limited to a certain number $hours/day$, which is determined by the designer to minimize battery degradation. It performs a $N$ steps closed loop simulation of the plant; while utilizing the same information as the MPC controller. The refrigerator power supply control command ($u_{fr}(k)$) for the current time step is computed through the thermostat logic. Moreover, it computes the total house load demand serviced ($E_{\text{\emph{total demand serviced}}}$), with which the energy mismatch ($E_{Mis} = E_{\text{\emph{total demand serviced}}}-E_{\text{\emph{total demand desired}}}$) is computed. $E_{Mis}$ is utilized to make the control decision for secondary load as described in Section~ \ref{subsubsection:IntelligentBaseline2}. 

\subsubsection{Secondary Load Logic Controller}\label{subsubsection:IntelligentBaseline2}
%Eq.~\eqref{eq:IntelligentBaseline_3} describes the rule based logic for generating the secondary load on-off command. It turns the secondary load on when there is no mismatch in total energy generated and the total house load desired demand as computed from the $N$ time steps simulation model. Moreover, when mismatch is present (i.e. generation is lower than the desired total house load) it turns the secondary load on in such a way that the mismatch is shedded through the secondary load, where $E_{SecTotal}$ is the total secondary load desired for the $N$ steps, $E_{SecProvided}$ is the total secondary load serviced within the $N$ steps, and both are reinitialized every $N steps$.
%\begin{align}
%u_{s}(k)= 
%\begin{cases}
%1,& \text{if }  E_{Mis} \geq 0 \; , E_{s}(k)>0\\
%1,& \text{if } E_{Mis}<0 \; , |E_{Mis}| \leq E_{SecTotal}, E_{s}(k)>0 \; , \\
%&  E_{SecProvided} \leq E_{SecTotal}-|E_{Mis}| \\
%0,& \text{otherwise, }
%\end{cases} \label{eq:IntelligentBaseline_3}
%\end{align}
 It turns the secondary load on ($u_{s}(k)=1$) when there is no mismatch ($E_{Mis}(k)=0$) as computed from the $N$ time steps simulation model. Moreover, when mismatch is present ($E_{Mis}(k)<0$) (i.e. generation is lower than the desired total house load) it turns the secondary load on in such a way that the mismatch is shedded through the secondary load alone. Finally, when mismatch is larger than that can be shed through the secondary load, it turns the secondary load off ($u_{s}(k)=0$).

\subsubsection{Battery Logic Controller}\label{subsubsection:IntelligentBaseline3}
%Once $u_{fr}(k)$ and $u_{s}(k)$ are decided as per previous sections, total house load ($E_{hl}$) can be computed as $E_{hl}(k)=u_fr(k)E_{fr}+u_s(k)E_{s}(k)$ and the battery control commands ($c$) and ($d$) can be computed using the baseline battery controller logic given in Section~\ref{subsubsection:Baseline2}, while fast charging command ($x_{bat}$) is computed using the rule based logic in Eq.~\eqref{eq:Battery4}.
%\begin{align}
%x_{bat}(k)= 
%\begin{cases}
%	2,& \text{if } c(k)=1 \; , E_{pv}(k)-E_{hl}(k) > \bar E_{bat}^{c} \\ 
%	1,& \text{if } c(k)=1 \; , E_{pv}(k)-E_{hl}(k) \leq \bar E_{bat}^{c} \\ 
%	0,& \text{otherwise, }
%\end{cases}  \label{eq:Battery4} 
%\end{align}
Once $u_{fr}(k)$ and $u_{s}(k)$ are decided as per Section~ \ref{subsubsection:IntelligentBaseline1} and Section~ \ref{subsubsection:IntelligentBaseline2} respectively, total house load ($E_{hl}$) is computed as $E_{hl}(k)=u_{fr}(k)E_{fr}+u_s(k)E_{s}(k)$ and the battery control commands ($c$) and ($d$) are computed using the baseline battery controller logic given in Section~\ref{subsection:Baseline}. While fast charging command ($x_{bat}$) is computed by augmenting this logic, where fast charging is commanded when excess energy from the PV allows for it i.e. $E_{pv}(k)-E_{hl}(k) > \bar E_{bat}^{c}$.

\section{Simulation Study Setup}\label{section:SRD}
The period selected for simulation is the time hurricane Irma passed over Gainesville, FL, USA, starting from its landfall on Sept. 11, 2017, to Sept. 17, 2017. Weather data is obtained from National Solar Radiation Database (\url{nsrdb.nrel.gov}). The simulations are run for 7 days starting at 00:00~hours (midnight) at day 1 (September 11, 2017)  with a planning horizon of 3 hours and a time step of 10 minutes ($\Delta t_{s}=10 \text{ mins}$, $N=18$) with battery initial state at $\bar E_{bat}$ (i.e., $E_{bat}(0)=\bar E_{bat}$) and the refrigerator initial temperature at 2$^{\circ}C$ (i.e., $T_{fr}(0)=2^{\circ}C$). 

%In addition, simulations have been carried out to see the effects of different: system sizes (on MPC, Baseline and Rule-Based controllers), house temperature models (on MPC controller), fast charging hours (on Rule-Based controller), and planning horizons (on MPC and Rule-Based controllers).

\subsection{Simulation Parameters}\label{subsection:SystemSizing}
\begin{figure}[htpb]
	\centering
	\includegraphics[scale=0.5]{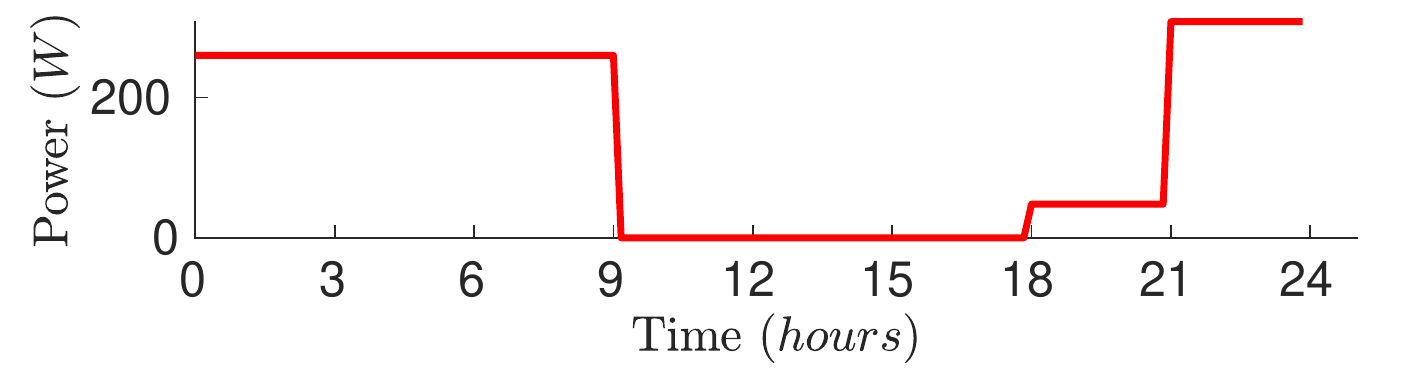}
	\caption{Secondary load demand (daily trajectory).}
	\label{fig:NCTrajectory}
\end{figure}

The house described in \cite{CuiHybrid:2019} consists of four bedrooms, a living room, and a kitchen. Hence, during and post hurricane period when power from grid is not available, the minimum load which will provide habitable conditions was decided to be an LED light for each room, a fan for each bedroom, and one refrigerator in the kitchen. Fig.~\ref{fig:NCTrajectory} illustrates the secondary load trajectory for a given day which is composed of: LED lights being on from 18:00~hours to 00:00~hours and fans running from 21:00~hours to 09:00~hours.
 
 We selected the Canadian Solar CS6K-285 polycrystalline panel ($\$100/panel$), and Trojan SPRE 12 225 (lead acid type) solar battery unit ($\$400/unit$). Lead acid battery is selected over Lithium-Ion (Li-ion) battery  despite the latter having performance advantages over the former in order to reduce cost, since Li-ion batteries are four times more expensive than lead acid batteries per $kWh$~\cite{DioufThe:2019}. The system was sized to service the total house load of the refrigerator, fans and lights for one day, which yielded a system consisting of 3 PV panels ($855W$) connected in parallel, 2 units of battery ($5400Wh$) connected in series. 

The internal house temperature, $T_{house}(k)$, for the plant is computed using the linear $4^{th}$ order ODE model given by~\cite{CuiHybrid:2019}, which models a typical, detached, two-story house in the USA.

%\textcolor{red}{\subsection{Simulation Parameters}}\label{subsection:SimulationParameters}
The parameters for the plant components; PV panels: $N_{pv}=3$ and $P_{pv}^{rated}=285$ $W$, $G_{std}=1000$ $W/m^{2}$; Battery: $\ubar E_{bat}=1080$ $Wh$, $\bar E_{bat}=5400$ $Wh$, $\bar E_{bat}^{c}=810$ $Wh$, $\bar E_{bat}^{dc}=844.5$ $Wh$, $\eta^{c}_{bat}=0.9$ and $\eta_{bat}^{dc}=0.9$ and Loads: Refrigerator - $P_{fr}^{rated}=250$ $W$, $\ubar T_{fr}=0$ $^{\circ}C$, $\bar T_{fr}=4$ $^{\circ}C$, Lights - $N_{l}=6$, $P_{l}^{rated}=8$ $W$ Fans - $N_{f}=4$, $P_{f}^{rated}=65$ $W$. The inverter efficiency is $\eta_{inv}=0.9$.

The parameters for the refrigerator thermal model are $C_{fr}=8.9374\times10^{3}$ $J/^{\circ}C$, $R_{fr}=1.4749$ $^{\circ}C/W$ and $COP=0.2324$. 

The parameters for the MPC are $\lambda_{1}=1$, $\lambda_{2}=1$, $\lambda_{3}=1$, $\lambda_{4}=10$, $\eta_{bat}^{c,dc,con}=1$, $\ubar\Gamma=-1$ and $\bar\Gamma=2$.

\subsection{Computation}\label{subsection:Computation}
The plant is simulated in MATLAB. The optimization problem is solved using GUROBI~\cite{gurobi:2019}, a mixed integer linear programming solver, on a Desktop Linux computer with 8GB RAM and a 3.60 GHz $\times$ 8 CPU.

\section{Results and Discussion}\label{section:ResultsDiscussions}
%In order to analyze the results we define two terms to describe the system state: Active System - when the system has enough power to maintain the refrigerator temperature within the prescribed bounds and Inactive System - when the system cannot produce power to maintain the refrigerator temperature within the prescribed bounds.

\subsection{Performance Metrics for Controller Comparison: }\label{subsection:ResultsDiscussions}
The performance of a controller is quantified by how well the primary and secondary loads were serviced during an extended outage. Hence, two metrics have been designed to compare the performance of the controllers. The Primary Resiliency Metric (PRM) is defined as
\begin{align}
\text{\emph{PRM}}= 1 - \dfrac{\int_{0}^{T_{sim}} \mathds{1}_{T_{fr}(t) > \bar T_{fr} +2} \,dt}{T_{sim}},  \label{eq:Resiliency1}
\end{align}
where $T_{sim}$ is the total simulation time, expressed in hours/day. It is a complement of the average hours per day the refrigerator temperature was above the tolerable upper limit.  The Secondary Resiliency Metric (SRM) is defined as
\begin{align}
\text{\emph{SRM}}= \dfrac{\int_{0}^{T_{sim}} \mathds{1}_{u_{s}(t).E_{s}(t) = E_{s}(t)} \,dt}{T_{sim}},  \label{eq:Resiliency2}
\end{align}
it is the percentage time secondary load was serviced compared to the desired secondary load trajectory. For both metrics, higher value means better performance.
\subsection{Performance Comparison of Controllers: }\label{subsection:ResultsDiscussions1}

\begin{figure*}[!t]
	\centering
	\begin{subfigure}[t]{0.32\textwidth}
		\centering
		\includegraphics[scale=0.40]{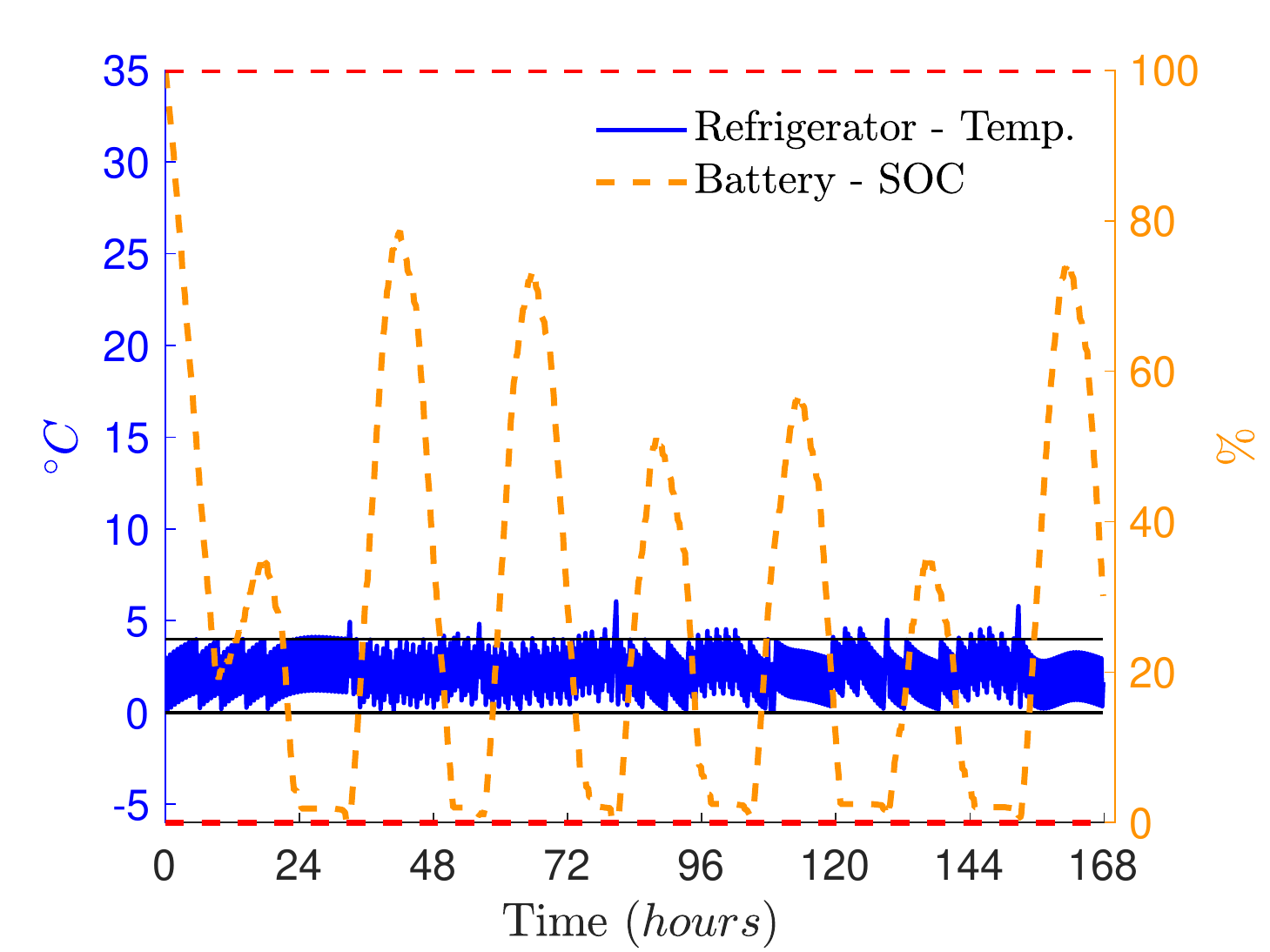}
		\caption{MPC: $T_{fr}$ and SoC.}
		\label{fig:Result_1}
	\end{subfigure}
	\begin{subfigure}[t]{0.32\textwidth}
		\centering
		\includegraphics[scale=0.40]{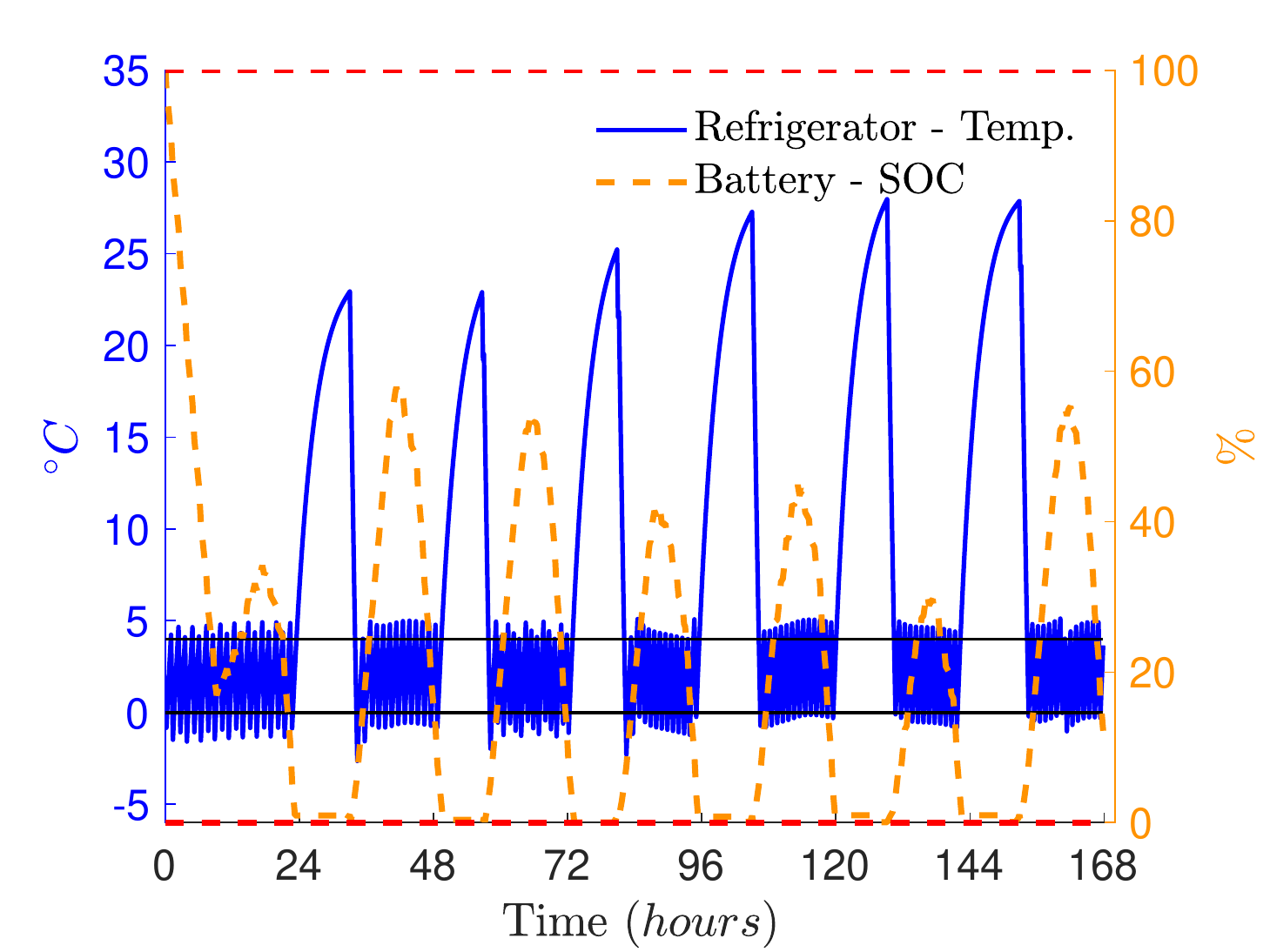}
		\caption{Baseline: $T_{fr}$ and SoC.}
		\label{fig:Result_2}
	\end{subfigure}
	\begin{subfigure}[t]{0.32\textwidth}
	\centering
	\includegraphics[scale=0.40]{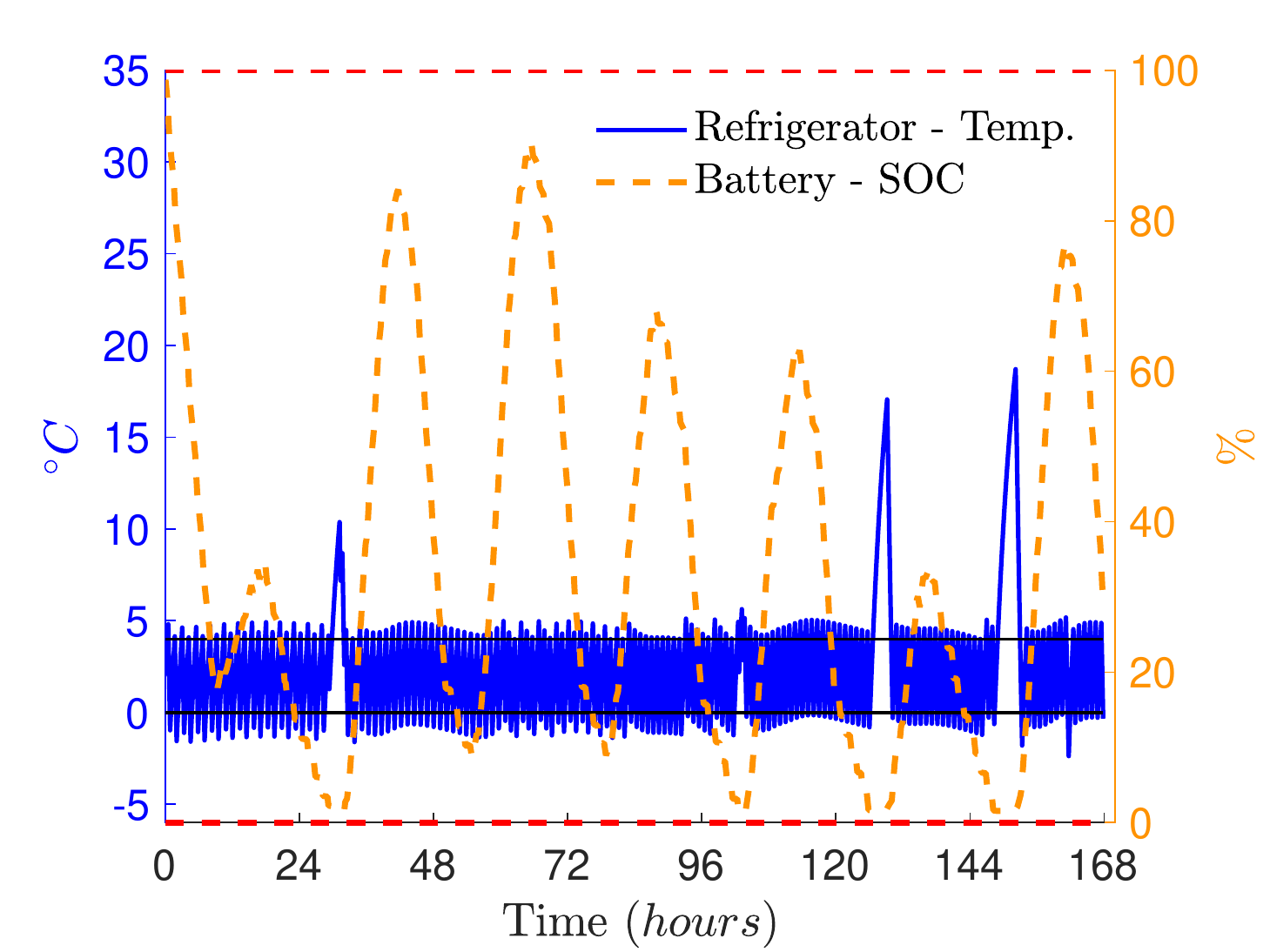}
	\caption{Rule-Based: $T_{fr}$ and SoC.}
	\label{fig:Result_3}
	\end{subfigure}
	
	\begin{subfigure}[t]{0.32\textwidth}
		\centering
		\includegraphics[scale=0.40]{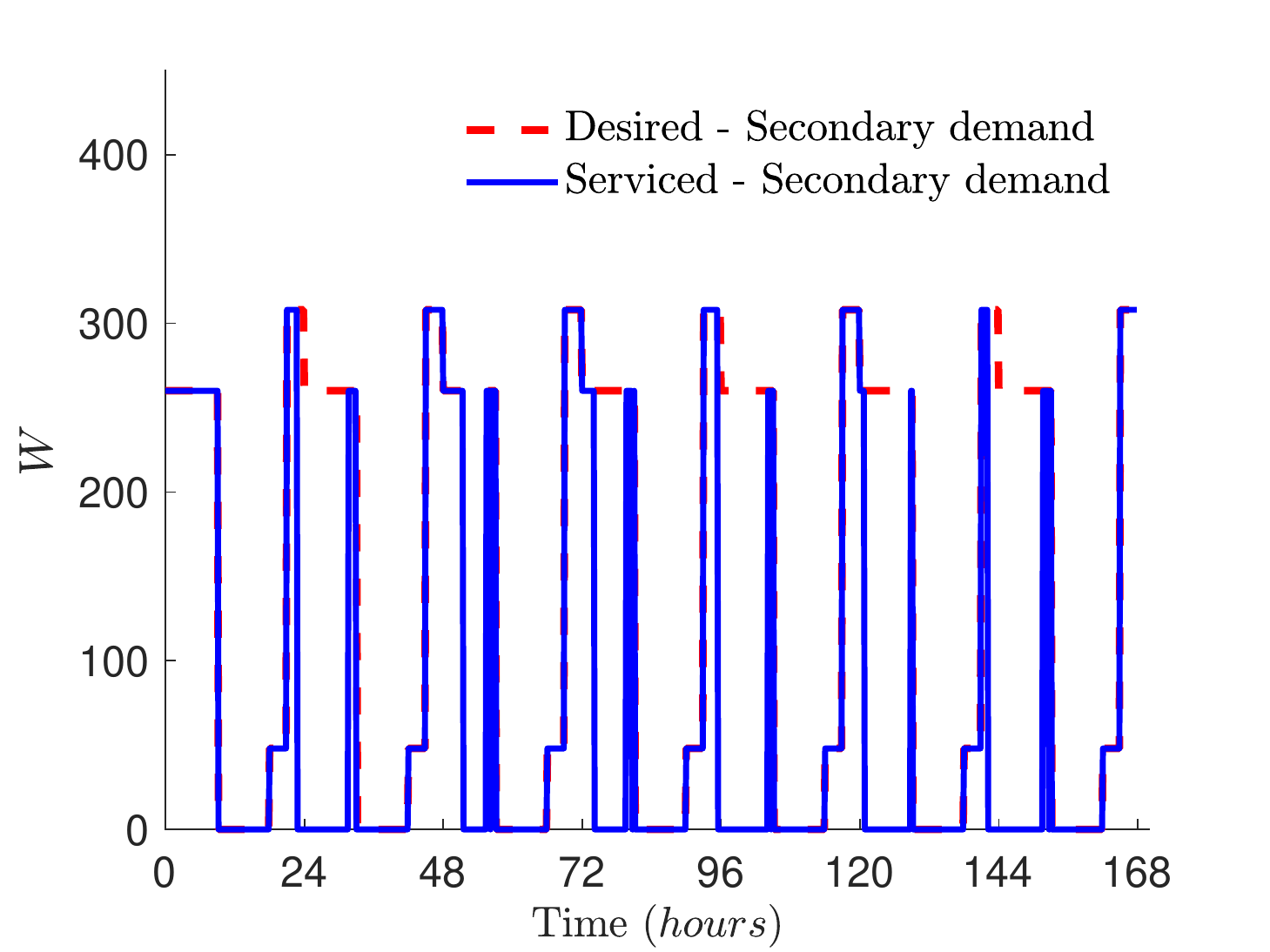}
		\caption{MPC: secondary performance.}
		\label{fig:Result_4}
	\end{subfigure} 
	\begin{subfigure}[t]{0.32\textwidth}
		\centering
		\includegraphics[scale=0.40]{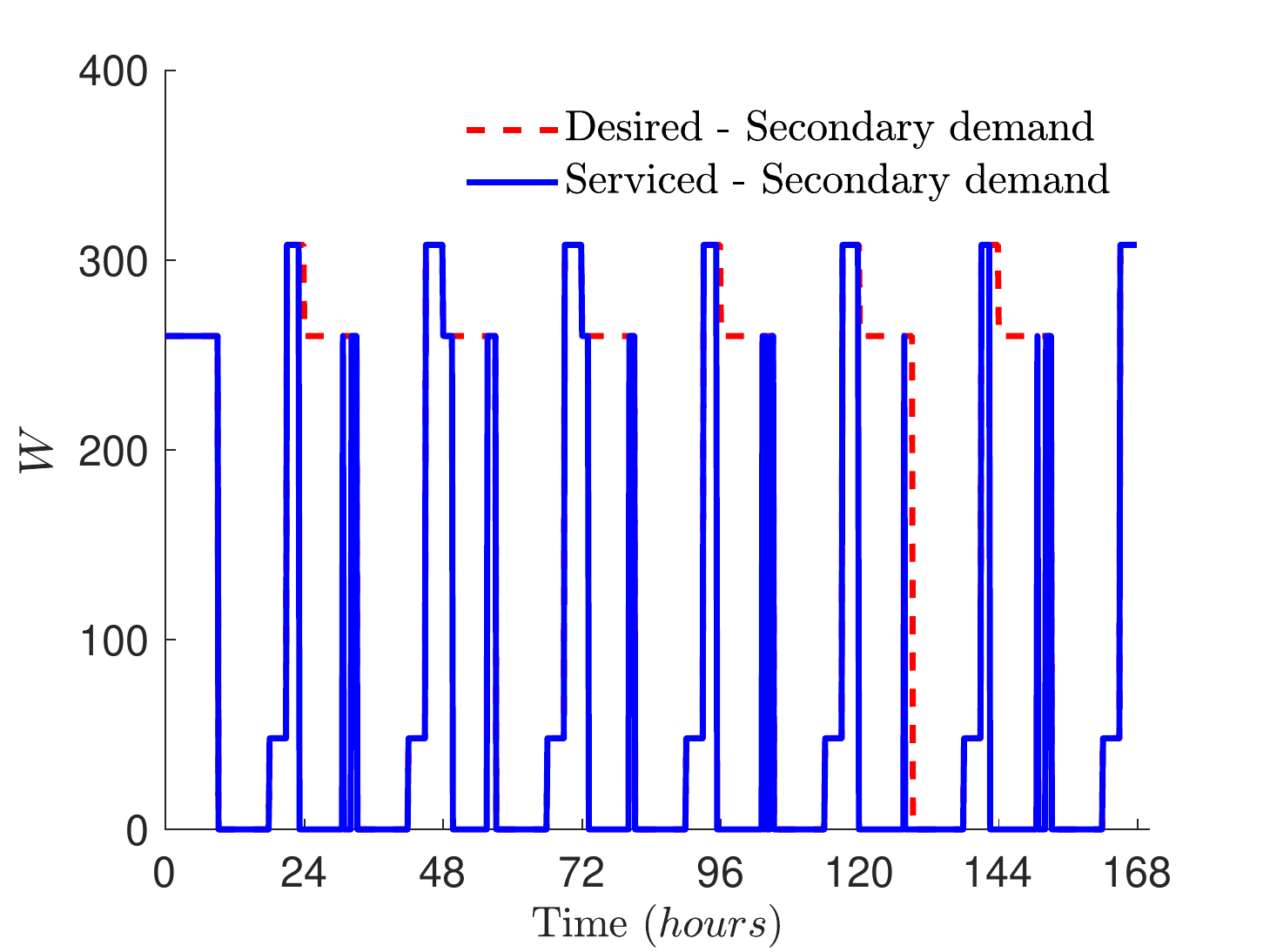}
		\caption{Baseline: secondary performance.}
		\label{fig:Result_5}
	\end{subfigure}
	\begin{subfigure}[t]{0.24\textwidth}
	\centering
	\includegraphics[scale=0.40]{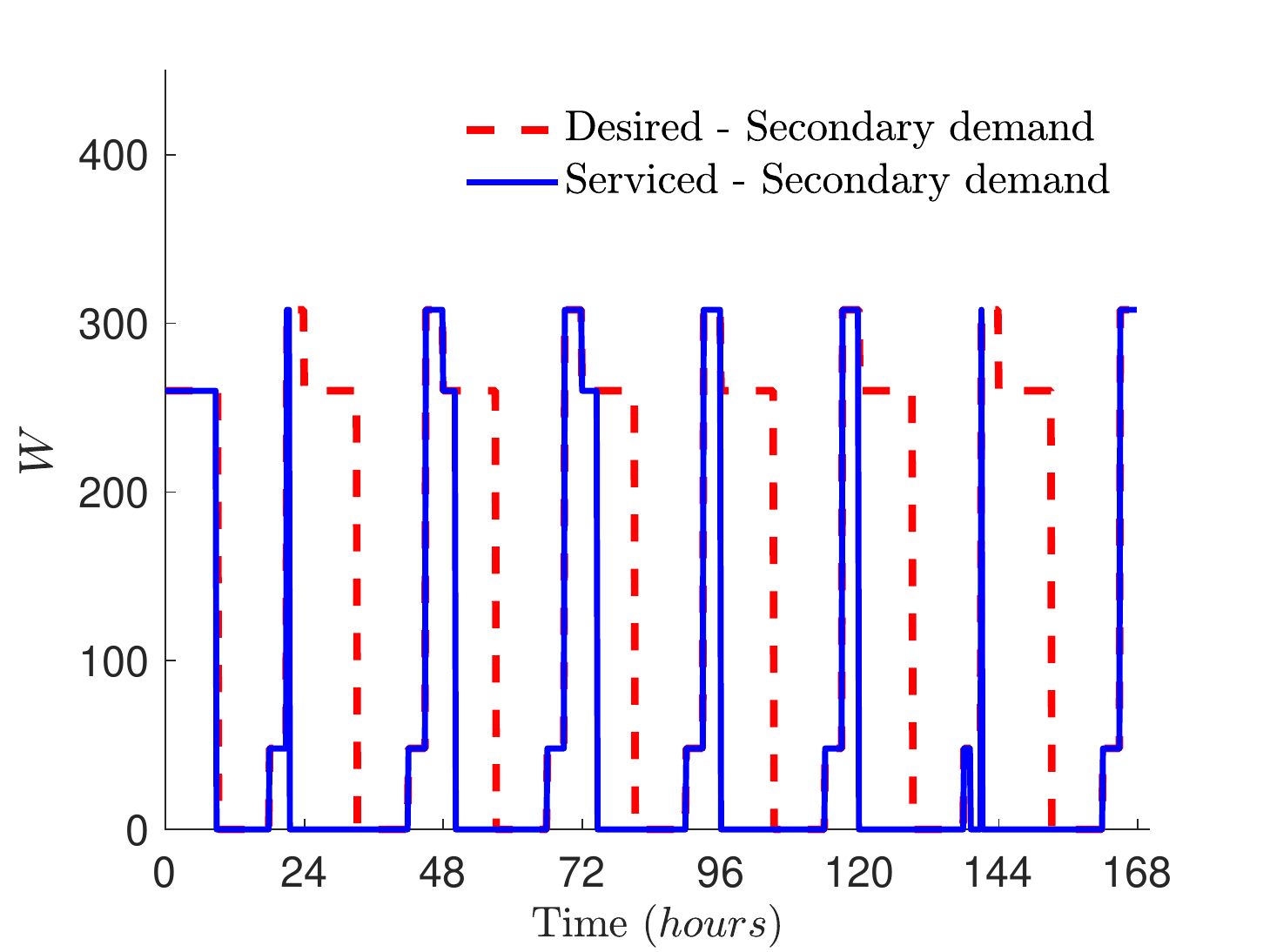}
	\caption{Rule-Based: secondary performance.}
	\label{fig:Result_6}
\end{subfigure}
	\caption{ Comparison of MPC, Baseline and Rule-Based controllers' performances for the week after hurricane Irma (Sept. 2017) in Gainesville, FL.}
	\label{fig:Results_PC}
\end{figure*}

Figure~\ref{fig:Results_PC} shows the simulation results when using the MPC, baseline and the Rule-Based controllers.
The MPC and Rule-Based controllers keep the refrigerator temperature within the prescribed limits for the entire 7 days with minor excursions; see Figures~\ref{fig:Result_1} and ~\ref{fig:Result_3}. In contrast, the baseline controller fails to do so for long periods; see Figure~\ref{fig:Result_2}.The PRM of baseline controller is  15.64~hours/day, while it is well over 20~hours/day for the MPC and Rule-Based controllers; see Table~\ref{tab:Results1}. The Centers for Disease Control and Prevention state that perishable foods (including meat, poultry, fish, eggs and leftovers) in the refrigerator should be thrown away if the power has been off for 4 hours or more~\cite{CdcKeep:2019}. Meaning, a PRM of at least 20~hours/day is needed. Thus, while the MPC and Rule-Based controllers will be able to keep perishable foods fresh for the entire seven days of the outage, with the baseline controller, the stored food will get spoiled after the very first day without grid power; of the three controllers MPC has the best PRM. 

\begin{table}[htbp]
	\caption{Performance comparison of MPC, baseline and Rule-Based controllers.}
	\label{tab:Results1}
	\begin{center}
		\begin{tabular}{|c||c||c|}
			\hline
			Controller Type & PRM $(h/day)$ & SRM $(\%)$  \\
			\hline
			MPC & 24 & 57.69 \\
			\hline
			Baseline & 15.64  & 47.94 \\
			\hline	
			Rule-Based & 22.04 & 48.21 \\
			\hline					
		\end{tabular}
	\end{center}
\end{table} 

Figures~\ref{fig:Result_4},~\ref{fig:Result_5} and~\ref{fig:Result_6}  show the trajectories of the secondary loads serviced by the baseline and MPC controllers respectively. It can be seen that none of the controllers are able to meet the secondary loads for the desired duration. However, the SRM of MPC controller is better (10\%) than the other two controllers, while the baseline and Rule-Based controllers perform similarly; see Table~\ref{tab:Results1}.

It is important to note that the SRM is poorest for the Rule-Based controller as the refrigerator on-off decision is made using the thermostat control and all other decisions are taken based on it i.e. the primary load is always favored over the secondary load. 

Hence, the MPC controller demonstrates superior performance in servicing both primary and secondary loads as compared to the baseline and Rule-Based controllers. The superior performance of the MPC controller is attributed to (i) its taking into account forecasts of disturbances (solar energy available, and desired trajectory for the secondary loads) in making decisions and (ii) making the trade off between various conflicting requirements by solving an optimization problem. While the baseline controller operates with the information consisting of just the present states ($T_{fr}(k)$ and $E_{bat}(k)$) of the system and its decision making is simple (rule based). And, even with equivalent information (sensing, forecasts and plant model) as the MPC controller, the  Rule-Based controller performs just slightly poorly than MPC controller in terms of PRM, while performing poorly in terms of SRM. This poor performance can be attributed to simple rule based decision making logic; which even though has access to same information as the MPC controller, is inadequate to make intelligent decisions to achieve the control goals. As the ability to make excellent trade-offs involving conflicting requirements is almost impossible with rule based logic in this problem, which makes the MPC control framework a rational choice for this problem.

\subsection{Effect of System Size on Controllers Performance:}\label{subsection:ResultsDiscussions6}
Table~\ref{tab:SystemSize} lists the various PV+battery sizes for which simulations were conducted. %Fig~\ref{fig:Result_BaselineMILP_ResiliencyPerformance_Size} shows the PRM, for different system sizes when using the three controllers. 
\begin{table}[htbp]
	\caption{System size, and description}
	\label{tab:SystemSize}
	\begin{center}
		\begin{tabular}{|c||c|}
			\hline
			System size& Description \\ 
			\hline
			A & 3 PV panels + 2 Battery units \\ 
			\hline
			B &  3 PV panels + 4 Battery units \\ 			 
			\hline
			C & 4 PV panels + 2 Battery units \\
			\hline
			D & 4 PV panels + 4 Battery units \\ 
			\hline
			E & 5 PV panels + 4 Battery units  \\
			\hline
			F & 6 PV panels + 4 Battery units  \\
			\hline			
		\end{tabular}
	\end{center}
\end{table}

\begin{figure}[htpb]
	\centering
	\includegraphics[scale=0.6]{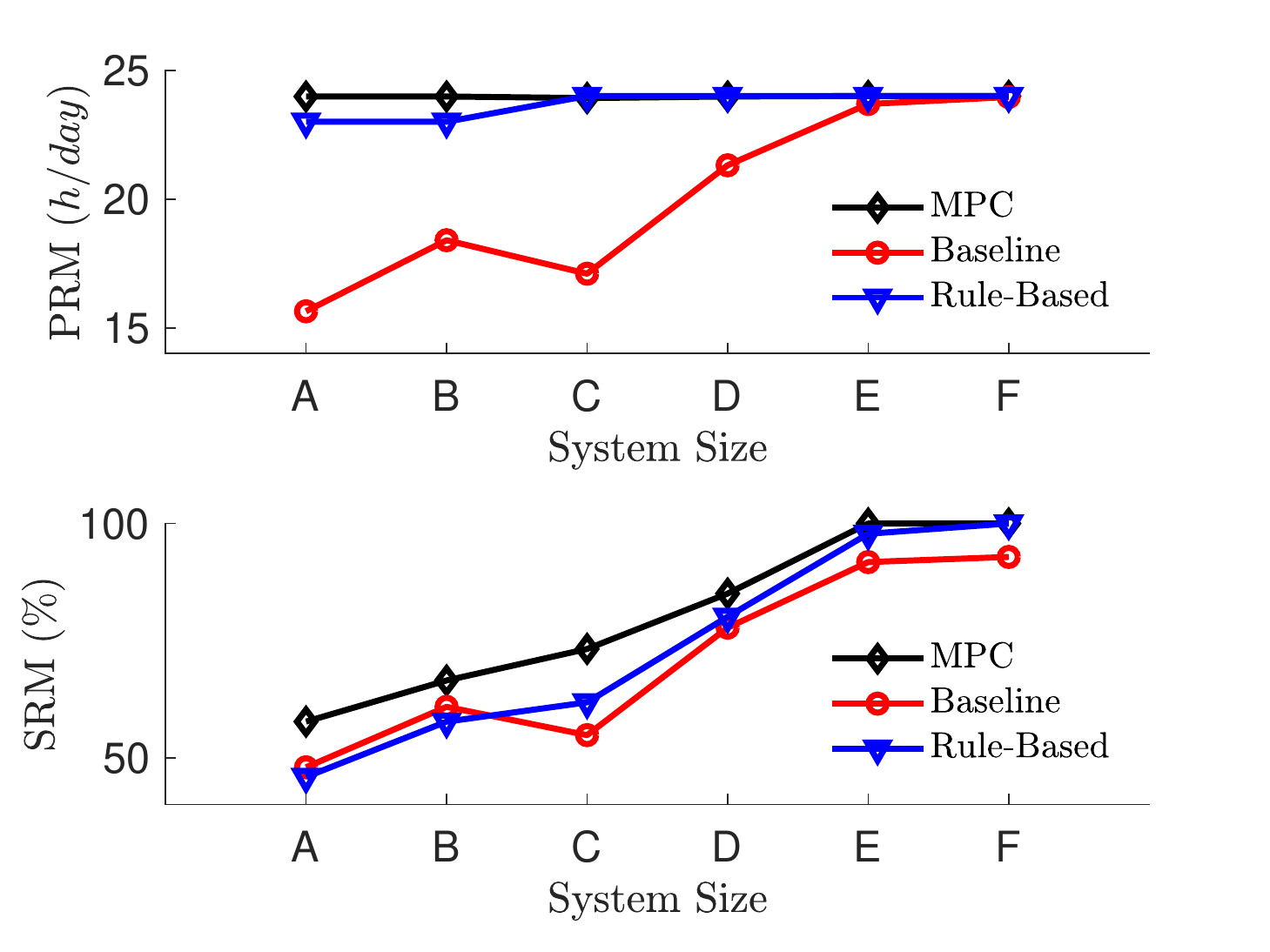}
	\caption{Resiliency performance metric of MPC, baseline and Rule-Based controllers with system size.} 
	\label{fig:Result_BaselineMILP_ResiliencyPerformance_Size}
\end{figure}
%\begin{table}[htpb]
%	\caption{\textcolor{red}{Performance comparison of MPC, baseline and Rule-Based controller with system sizes.}}
%	\label{tab:Results2}
%	\begin{center}
%		\begin{tabular}{| *{6}{c||}c|}
%			\hline
%			 Controller & \multicolumn{2}{c||}{MPC} & \multicolumn{2}{c||}{Baseline} & \multicolumn{2}{c|}{Rule-Based}\\
%			\hline
%			System Size & PRM & SRM & PRM & SRM & PRM & SRM\\			
%			\hline
%			A & 0.02  & 57.69    & 8.36  & 47.94    & 1  & 45.88   \\
%			\hline
%			B & 0.02 & 66.48  & 5.6 & 60.85  & 1 & 57.69  \\
%			\hline
%			C & 0.08 & 73.21 &  6.9 & 54.81 &  0 & 61.81  \\
%			\hline
%			D & 0.02 & 85.03  & 2.69 & 77.75  & 0 & 80.08  \\
%			\hline
%			E & 0 & 100  & 0.31 & 91.76  & 0 & 97.8  \\
%			\hline	
%			F & 0 & 100  & 0.04 & 92.86  & 0 & 100  \\
%			\hline		
%		\end{tabular}
%	\end{center}
%\end{table} 

Fig~\ref{fig:Result_BaselineMILP_ResiliencyPerformance_Size} shows that for the baseline controller to achieve a similar PRM as MPC controller, the system cost/size has to be doubled. It should be noted that the Rule-Based controller's PRM is slightly lower as compared to MPC controller's PRM at smaller system sizes; however, it achieves a similar PRM performance as the MPC controller with a $1/5^{th}$ increase in system cost/size demonstrating its edge over the baseline controller. It can be noticed that SRM is directly proportional to the system size. SRM of MPC controller is always higher than that of the baseline (9.41\% on average), while it is higher than that of the Rule-Based controller (6.53\% on average) on all but the biggest system size where they are equal. Hence, the equipment cost required to achieve a similar level of resiliency performance can be halved by the MPC controller as compared to the baseline controller. 

\ifArxivx

\subsection{Effect of House Temperature Parametric Model on MPC Controller: }\label{subsection:ResultsDiscussions2}
It is observed that the resiliency performance of the MPC controller is unaffected by the choice of house temperature model; see Table~\ref{tab:Results3}. This demonstrates that the simple parametric house temperature model developed in (\ref{eq:HouseThermalParametricModel}) does not deteriorate the performance of the MPC controller as compared to the complex linear $4^{th}$ order ODE thermal model of the house given in~\cite{CuiHybrid:2019}. Moreover, it helps us avoid this complex thermal model of the house to be part of the MPC controller. This is desirable as this thermal model is designed for a typical detached two-story house in the USA and would have required the additional effort of model estimation for any other house.  

\begin{table}[htpb]
	\caption{Performance comparison of MPC controller with house temperature models.}
	\label{tab:Results3}
	\begin{center}
		\begin{tabular}{| *{2}{c||}c|}
			\hline
			House temperature model & PRM ($h/day$) & SRM ($\%$)\\			
			\hline
			Parametric & 24 & 57.69   \\
			\hline
			Linear $4^{th}$ order ODE & 24 & 57.42  \\
			\hline		
		\end{tabular}
	\end{center}
\end{table} 

\subsection{Effect of Fast Charging Hours on Rule-Based Controller: }\label{subsection:ResultsDiscussions3}
For the MPC controller, fast charging is limited (which limits battery degradation) through cost function design. On the other hand in the Rule-Based controller this is done by setting the maximum allowed fast charging hours for a given day. It can be seen that the PRM of Rule-Based controller is not affected much by increasing the maximum allowed fast charging hours, while the SRM generally increases slightly and plateaus at 5 hours of fast charging per day; see Table~\ref{tab:Results4}.  
\begin{table}[htpb]
	\caption{Performance of Rule-Based controller with fast charging hours.}
	\label{tab:Results4}
	\begin{center}
		\begin{tabular}{| *{2}{c||}c|}
			\hline
			Fast charging hours ($hours$) & PRM ($h/day$) & SRM ($\%$) \\			
			\hline
			1 & 22.31 & 41.21  \\
			\hline
			2 & 23 & 43.27  \\
			\hline
			3 & 22.98 & 45.19  \\
			\hline		
			4 & 22.73 & 46.02  \\
			\hline
			5 & 23 & 45.88  \\
			\hline
			6 & 23  & 45.88   \\
			\hline	
		\end{tabular}
	\end{center}
\end{table}

\subsection{Effect of Planning Horizon on MPC and Rule-Based Controllers:}\label{subsection:ResultsDiscussions4}

Generally, longer the planning horizon larger the future information data is provided to the controller, and better the performance becomes. However, the computational cost increases with planning horizon for solving the underlying optimization problem in case of MPC based controllers. 

\begin{table}[htpb]
	\caption{Performance comparison of MPC and Rule-Based controller with planning horizons.}
	\label{tab:Results5}
	\begin{center}
		\begin{tabular}{| *{4}{c||}c|}
			\hline
			Controller & \multicolumn{2}{c||}{MPC} & \multicolumn{2}{c|}{Rule-Based}\\
			\hline
			PH ($hours$) & PRM & SRM  & PRM & SRM \\			
			\hline
			1 & 13.08 & 17.86  & 19.75 & 50.41 \\
			\hline
			3 & 24 & 57.69  & 22.04 & 48.21  \\
			\hline
			6 & 23.96 & 56.32  & 22.98 & 47.12  \\
			\hline
			12 & 23.79 & 52.75  & 22.98 & 47.12  \\
			\hline
			24 & 23.75  & 53.02   & 23.81  & 37.77  \\
			\hline			
		\end{tabular}
	\end{center}
\end{table} 

It is observed from the Table~\ref{tab:Results5} that the performance of the MPC controller is poorest with the smallest planning horizon (1 hour) indicating that future information is necessary for better performance, and performance peaks at 3 hours. However, further increase in planning horizon deteriorates the performance owing to the GUROBI MILP solver not being successful in converging to a solution 100\% times but rather stalling, see Table~\ref{tab:Results6} leading to utilization of a sub-optimal solution.

\begin{table}[htpb]
	\caption{Computational performance of MPC controller with planning horizons.}
	\label{tab:Results6}
	\begin{center}
		\begin{tabular}{| *{2}{c||}c|}
			\hline
			Controller & \multicolumn{2}{c|}{MPC} \\
			\hline
			PH ($hours$) & Average solver time ($sec$) & Solver success ($\%$)   \\			
			\hline
			1 & 0.02 & 100   \\
			\hline
			3 & 0.06 & 100    \\
			\hline
			6 & 0.21 & 100    \\
			\hline
			12 & 5.97 & 97.83    \\
			\hline
			24 & 60  & 91.49    \\
			\hline			
		\end{tabular}
	\end{center}
\end{table}

On the other hand, it can be observed that the PRM of the Rule-Based controller improves with planning horizon, while its SRM decreases; see Table~\ref{tab:Results6}. This demonstrates that the Rule-Based controller is able to use the additional future information well for the primary load, but is not intelligent enough to use it effectively for the secondary load.

\else

The following sensitivity studies have been conducted to evaluate the proposed controllers:
\begin{enumerate}
\item Effect of temperature model on the MPC controller's performance, by comparing with the performance of the mPC controller proposed in earlier work~\cite{GaikwadSmartCCTA:2020} which used a 4th order dynamic model of the house temperature.
	\item Effect of fast charging hours on the Rule-Based controller's performance.
	\item Effect of planning horizon on both MPC and Rule-Based controller.
\end{enumerate}
The details are omitted here due to lack of space; the interested reader can find them in~\cite{GaikwadIncreasing_ArXiV:2021}. The summary of the results are as such. The MPC controller's performance (both PRM and SRM) is not improved by the more accurate and complex $4^{th}$ order ODE model. Second, increasing the allowed fast charging hours per day of the Rule-Based controller does not affect its PRM significantly, however it improves its SRM. Third, increasing the planning horizon deteriorates the MPC controller's performance slightly due to optimization solver's failures for large dimensional problem. On the other hand for the Rule-Based controller the PRM improves while its SRM deteriorates. 
\fi

\section{Conclusion}\label{section:Conclusions}
The study provides further verification of the premise that intelligence can reduce cost of energy resiliency to hurricane induced blackouts. Our preliminary work~\cite{GaikwadSmartCCTA:2020} showed that intelligent controller using MPC performs better than a simple baseline controller that simply supplies demand until energy runs out. That was not surprising, but it was not clear whether similar performance could be obtained by a little more intelligent decision making instead of using MPC if it too can avail of the sensing and forecast information that MPC uses. The results here shows that such an intelligent Rule-Based controller requires a $20\%$ increase in system size, while the baseline controller requires a $100\%$ increase in system size, to provide the same level of resiliency performance as the proposed MPC controller. For a fixed PV-battery size, the MPC controller moderately outperforms the Rule-Based controller in servicing the primary load (refrigerator) and significantly outperforms it in servicing the secondary load. This observation indicates that intelligent and automated decision making that is rooted in real-time optimization provides better performance compared to complex rule based logic even when the two are provided the same sensing and forecast information. The sensitivity studies conducted provide further confidence on the performance of the MPC controller.

This study opens up many directions for future research. These include analysis of sensitivity to forecast errors, reducing the information requirements (both sensing and forecasts) of the control algorithm, optimal sizing of a PV-battery system taking into account resiliency during disasters and energy savings during normal times. 

\bibliographystyle{IEEEtran}
\bibliography{\DiCEbibPATH/resiliency,\DiCEbibPATH/disaster,\DiCEbibPATH/Barooah,\DiCEbibPATH/optimization,\DiCEbibPATH/grid,\DiCEbibPATH/building,\DiCEbibPATH/systemid}

\end{document}